\documentclass[useAMS,referee,usenatbib]{biom}
\usepackage{amsmath}
\usepackage{graphicx}
\usepackage{subfigure}
\usepackage{indentfirst}
\usepackage{color,xcolor}
\usepackage{graphicx,psfrag,epsf}
\usepackage{enumerate}
\usepackage{amsfonts}
\usepackage{booktabs}
\usepackage{epstopdf}
\usepackage{threeparttable}

\title{Functional Mixed effects Model for Joint Analysis of Longitudinal and Cross-Sectional Growth Data}

\author{Long Chen, Ji Chen and Yingchun Zhou$^*$\email{yczhou@stat.ecnu.edu.cn}\\
Key Laboratory of Advanced Theory and Application in Statistics\\
and Data Science-MOE, School of Statistics, East China Normal University,\\
3663 North Zhongshan Road, Shanghai, P.R. China}

\begin{document}

%\date{{\it Received October} 2007. {\it Revised February} 2008.  {\it Accepted March} 2008.}

%\volume{64}
%\pubyear{2008}
%\artmonth{December}

%\doi{10.1111/j.1541-0420.2005.00454.x}

\label{firstpage}

\begin{abstract}
A new method is proposed to perform joint analysis of longitudinal and cross-sectional growth data. Clustering is first performed to group similar subjects in cross-sectional data to form a pseudo longitudinal data set, then the pseudo longitudinal data and real longitudinal data are combined and analyzed by using a functional mixed effects model. To account for the variational difference between pseudo and real longitudinal growth data, it is assumed that the covariance functions of the random effects and the variance functions of the measurement errors for pseudo and real longitudinal data can be different. Various simulation studies and real data analysis demonstrate the good performance of the method.
\end{abstract}


\maketitle

\section{Introduction}
In biomedical and other research fields, it often occurs that data from different sources and even of different types are obtained to study the same problem. A typical example is that when studying the influencing factors of a particular disease, both cohort studies and cross-sectional studies are performed and data are collected. Traditionally these two types of data are analyzed separately and there are well-developed methods for analyzing each type of data. Comprehensive presentations of longitudinal data analysis methods can be found in, for example, \citet{singer2003applied}, \citet{hedeker2006longitudinal} and \citet{fitzmaurice2012applied}, and those of the cross-sectional data analysis methods can be found in, for example, \citet{chatterjee2015regression} and \citet{Wooldridge2010}. However, there is little literature on how to combine these two data types to jointly analyze a problem.

Since cohort studies focus on observing a continuous phenomenon and often present the characteristics that change with time, functional data analysis is a natural tool for analyzing these types of data. Particularly, if one would like to combine cross-sectional data with cohort data, and the two data sets are often observed at different time points, functional data analysis which can easily deal with irregularly sampled data becomes a much more powerful tool than other methods.

Functional data analysis has become a popular research area in the past decade. \citet{ramsay2005functional} gave an overview of various models and many case studies about functional data. \citet{ferraty2006nonparametric} presented nonparametric statistical methods for functional data analysis. \citet{ramsay2007applied} illustrated how functional data analysis work out in practice in a diverse range of subject areas, including economics, education, archaeology, criminology, psychology, auxology, meteorology, biomechanics, etc. \citet{yao2005functional} proposed a nonparametric method to perform functional principal component analysis for the case of sparse longitudinal data. In the area of functional mixed effects model, \citet{guo2002functional} introduced a class of functional mixed effects models based on smoothing splines. \citet{morris2006wavelet} proposed a Bayesian wavelet-based method to fit functional mixed effects model. \citet{Chen2011} and \citet{chen2018functional} proposed penalized spline-based methods for functional mixed effects models with varying coefficients. %\citet{kauermann2011functional} presented a method for joint modeling of mean and variance in longitudinal data using penalized splines. which model both components simultaneously via rich spline bases.

In this article, we propose an effective method that combines longitudinal and cross-sectional growth data in the same analysis, so as to make full use of all collected data. To be specific, we first cluster cross-sectional data with only one observation for each individual into groups, and each group of observations can be regarded as repeated observations from a pseudo individual. Then we combine the pseudo longitudinal data with real longitudinal data and analyze the combined data set by using a functional mixed effects model. To account for the variational difference between pseudo and real longitudinal growth data, it is assumed that the covariance functions of the random effects and the variance functions of the measurement errors for pseudo and real longitudinal data can be different.

The proposed method distinguishes from the traditional parametric methods of functional mixed effects model \citep{diggle2002analysis} in that there is no need to assume the measurement error working independent with constant variance or having its covariance function specified by a parametric model. The method also distinguishes from the method proposed by \citet{Chen2011} in that the subject-specific random effects of pseudo longitudinal data and real longitudinal data may have different covariance functions, similarly the measurement errors of these two sets of data may have different variance functions. By accurately estimating the different covariance and variance functions, one gains more efficiency in estimating the fixed effects parameters and population mean function \citep{Fan2007Analysis}.

The nonparametric components of the proposed model are estimated by penalized spline (P-spline) which is a variant of smoothing spline with more flexible choices of bases, knots and penalties. Alternatively, P-spline can be viewed as least square regression spline with a roughness penalty. P-splines were originally proposed by \citet{O1986statistical} and has gained popularity since \citet{eilers1996flexible} and \citet{ruppert2003carroll}. Comprehensive overviews of the development of P-spline can be found in \citet{ruppert2009semiparametric} and \citet{eilers2015twenty}. It is proved that P-spline as a reduced-rank smoother can be asymptotically as effective as full-rank estimates obtained by smoothing splines \citep{li2008asymptotics,claeskens2009asymptotic}.

The rest of the paper is structured as follows: Section 2 introduces a practical example that motivated us to work on this problem. Section 3 describes the whole procedure of the proposed method. In Section 4, various simulations are conducted to investigate the performance of the proposed method. In Section 5, the method is applied to infant growth data to produce interesting results. Section 6 discusses several issues and prossible extensions of the method.

\section{A Motivating Example}

A practical problem that motivated us to work on the methodology is introduced here. The goal of the problem is to study the dynamic relationship between growth (e.g. in height) and gender for infants from 0 to 2 years old. There exist two sets of data: a longitudinal data set obtained from a cohort study and a cross-sectional data set obtained from a survey. Both data sets include sex, height, birth weight, parental heights, birth place and other variables of infants between 0 and 2 years old. Detailed information of the two data sets are as follows.

\textbf{Longitudinal Data}: 251 infants in Shanghai were followed from birth to 24 months, with measurements taken at 42 days, 3 months, 6 months, 12 months, 18 months and 24 months, respectively. Some of the observations were missing.

\textbf{Cross-Sectional Data}: Data of 1083 infants are collected from eight provinces in China. Each infant was observed at a random time point from birth to 24 months.\vspace{3mm}

\begin{figure}[h!]
	\centering
	\begin{minipage}[c]{0.36\textwidth}
		\centering
		\includegraphics[width=1\textwidth]{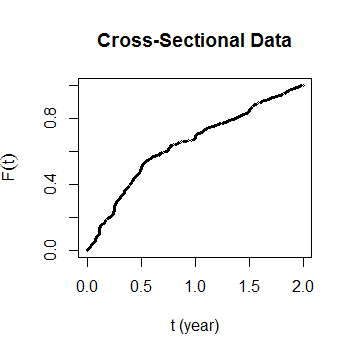}
	\end{minipage}
	\begin{minipage}[c]{0.36\textwidth}
		\centering
		\includegraphics[width=1\textwidth]{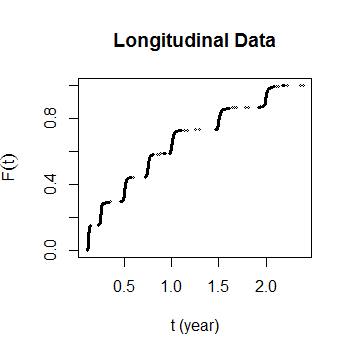}
	\end{minipage}
	\caption{The distributions of the observation time points of the cross-sectional data set and the longitudinal data set. \label{figure:timepoints} }
\end{figure}

The distributions of observation time points of the two data sets are shown in Figure \ref{figure:timepoints}. It can be seen that the time points that measurements were taken for the cross-sectional data are almost everywhere between 0 and 2 years. Even for the longitudinal data where time points were preset by investigators, the observation time points are still scattered, which makes it difficult to analyze using traditional methods.

In facing such data, traditional methods usually analyze the two data sets separately. However, it is preferred to maximize the sample size when analyzing the problem by making full use of all available data, so as to best estimate the dynamic relationship between growth and gender. This goal motivates us to develop a new methodology.

\section{Methodology}
The proposed method consists of two major steps. In Step 1, the observations in the cross-sectional data are clustered into groups by some clustering methods such as K-means clustering, and a pseudo longitudinal data set is formed. In Step 2, a combined functional mixed effects model is constructed, with the assumption that the covariance functions of the random effects and the variance functions of the errors are different for the pseudo and real longitudinal data sets.

In estimating the model, an iteration algorithm is adopted by iterating between Step 2A and Step 2B until convergence is reached. Step 2A: Given the difference between the covariance functions associated with random effects and the variance functions associated with the errors of the two data sets, obtain the estimates of the population mean function, fixed coefficients and time-varying coefficients; Step 2B: Given the population mean function, fixed coefficients and time-varying coefficients, obtain estimates of the difference between the covariance functions of the random effects and the variance functions of the errors of these two data sets, and hence covariance functions and variance functions of the two data sets.

\subsection{Clustering of the cross-sectional data (Step 1)}
It is important to properly cluster the observations in the cross-sectional data set to form a pseudo longitudinal data set. First one needs to identify the variables that can best group the individuals. It is preferred to choose those variables that have big influence on the response variable so that different patterns of the responses can be captured by using these variables. In the real data analysis part of the paper, we first applied a functional mixed effects model to the longitudinal data set to select statistically significant variables, and then cluster according to these variables.

There are many clustering methods for multivariate data. Here we choose K-means clusering method because it's fast and produce good results. Other methods that suit the data forms can be used as well.

Once the observations in the cross-sectional data are clustered, one can treat each cluster of observations as a pseudo individual with repeated measurements. Since the pseudo individuals consist of observations from different real individuals across different locations, the within-subject variation, the between-subject variation and the error variation are all likely to be larger than those of the real longitudinal data which consist of repeated observations of real individuals. Therefore one needs to consider different variance and covariance functions in the combined model.

\subsection{Functional mixed effects model for the combined data (Step 2)}
The functional mixed effects model for the combined data are constructed following \citet{Chen2011}, and innovatively take into account that the subject-specific random effects of the two data sets may have different covariance functions and the errors of the two may have different variance functions. The model is written as follows, where $i$ indexes subjects, $j$ indexes visits and $k$ indexes data sets:
\begin{equation}\label{g1}
y_{ij}^{(k)}=\mu(t_{ij})+(x_{ij}^{(k)})^T\alpha+\omega_{ij}^{(k)}\beta(t_{ij})
+\nu_{i}^{(k)}(t_{ij})+\epsilon^{(k)}_{ij}(t_{ij}),
\end{equation}
\[\nu_{i}^{(k)}(t)\sim W(0,\gamma_{k}),\ \epsilon_{i}^{(k)}\sim N(0,V_{k_{i}}^{\frac{1}{2}}R_{i}(\theta)V_{k_{i}}^{\frac{1}{2}}),\]
\[ V_{k_{i}}=diag\{\sigma_{k}^{2}(t_{i1}),\dots,\sigma_{k}^{2}(t_{i,m_i})\},\]
\[i=1,\dots,n_{k},\ j=1,\dots,m_{i},\ k=1,2,\ n=n_{1}+n_{2},\]
where $\mu(t)$ is population mean function, $x_{ij}$ is a vector of covariates with a fixed coefficient vector $\alpha$, $\omega_{ij}$ is a vector of covariates with time-varying coefficients $\beta(t_{ij})$, $\nu_{i}^{(k)}(t)$ are functional subject-specific random effects assumed to be independent among subjects, and follows a Gaussian process $W(0,\gamma_{k})$ with covariance function $\gamma_{k}(s,t)$. $\epsilon_i^{(k)}=(\epsilon_{i1}^{(k)},\dots,\epsilon_{im_i}^{(k)})^T$ is a vector of errors which is independent of the random effects, and $\epsilon_{ij}^{(k)}$ has a variance function $\sigma_{k}^{2}(t)$. $R_{i}(\theta)$
is a parametric correlation matrix such as AR-1 (first-order autoregressive) or compound symmetry
with $\theta$ being the vector of unknown parameters.

To estimate the model, suppose $\mu(t), \beta(t), \nu_{i}^{(k)}(t), log\,\sigma_{k}^{2}(t)$ can be approximated by
\[\mu(t)=B_{\mu}(t)\beta_{\mu},\qquad \beta(t)=B_{c}(t)\beta_{c},\]
\[\nu_{i}^{(k)}(t)=B_{\nu}(t)S_{k}\xi_{i},\qquad log\,\sigma_{k}^{2}(t)=B_{\sigma}(t)M_{k}\eta,\]
where$B_{\mu}(t)$, $B_{c}(t)$, $B_{\nu}(t)$, and $B_{\sigma}(t)$ are vectors of basis functions with possibly different orders or different numbers of knots; $\beta_ {\mu}$, $\beta_{c}$, $\xi_{i}$, and $\eta$ are their corresponding basis coefficients. $S_{k}$ and $M_{k}$ are matrices of parameters that respectively reflect the difference between the covariance functions of the random effects and the difference between the variance functions of the errors of the two data sets. Let
$B_{\nu}^{(k)}(t)=B_{\nu}(t)S_{k}$, $B_{\sigma}^{(k)}(t)=B_{\sigma}(t)M_{k}$, one obtains
\[\nu_{i}^{(k)}(t)=B_{\nu}^{(k)}(t)\xi_{i},\qquad log\,\sigma_{k}^{2}(t)=B_{\sigma}^{(k)}(t)\eta,\]
\[\gamma_{k}(s,t)=B_{\nu}^{(k)}(s)\Omega (B_{\nu}^{(k)}(t))^{T}, \qquad\text{where}\;\Omega
=\mathrm{cov}(\xi_{i}).\]

In selecting the spline basis function, we use the truncated polynomial basis function, which is written as:
\begin{equation}\label{g2}
m(t;a)=a_{0}+a_{1}(t)+a_{2}(t^2)+\dots+a_{p}
(t^p)+\sum_{n=1}^{N}a_{p+n}(t-knots_{n})_{+}^{p},
\end{equation}
where $p$ is the order of the basis function, $knots_{1}<knots_{2}<\dots<knots_{N}$ are $N$ fixed knots, $a=(a_{0},a_{1},\dots a_{p+N})$ is a vector of basis coefficients. With these, model (1) can be re-written as:
\begin{equation}\label{g3}
Y_ {i}^{(k)}=X_{i}^{(k)}\beta+Z_{i}^{(k)}\xi_{i}+\epsilon_{i}^{(k)},
\end{equation}
\[\xi_{i}\sim N(0,\Omega),\qquad \epsilon_{i}^{(k)}\sim N(0,V_{k_{i}}^{\frac{1}{2}}R_{i}V_{k_{i}}^{\frac{1}{2}}),\]
where
$Y_{i}^{(k)}=(y_{ij}^{(k)})_{j=1,\dots,m_{i}}$, $X_{i}^{(k)}=(x_{i}^{(k)},B_{\mu}^{i},B_{c}^{i})$,
$Z_{i}^{(k)}=((B_{\nu}^{(k)}(t_{i1}))^{T},\dots,(B_{\nu}^{(k)}(t_{im_{i}}))^{T})^{T}$,
$\beta=(\alpha^{T},\beta_{\mu}^{T},\beta_{c}^{T})^{T}$, and
$B_{c}^{i}=(\omega_{i1}B_{c}^{T}(t_{i1}),\dots,\omega_{im_{i}}B_{c}^{T}(t_{im_{i}})).$

Model (3) is similar to the usual multivariate linear mixed effects model, which is estimated through an iterative procedure between Step 2A and 2B.

\subsection*{Step 2A}
Given $S_k$ and $M_k$, the estimate of $\beta$ which includes the parameters in the population mean function $\mu(t)$, fixed coefficients $\alpha$ and time-varying coefficients $\beta(t)$ can be obtained.

During the first iteration, let $S_{k}=diag\{1,\dots,1\}$, $M_{k}=diag\{1,\dots,1\}$,
then the estimation method proposed by \citet{Chen2011} can be applied. The method is briefly described as follows. First the penalized joint log likelihood of $Y_{i}^{(k)}$ and $\xi_{i}^{(k)}$ is defined, then given the initial values of the variance components, one obtains the initial estimates of $\beta$ and $\xi_{i}$ by minimizing the penalized joint log likelihood function, and the estimates of the between-subject variance component $\Omega$ through restricted maximum likelihood. Then based on the above estimates, one adopts the EM Algorithm to update the above estimates. In the iteration, a Newton-Raphson based method is applied to estimate $\theta$ and $\eta$. And a likelihood-based selection approach is employed to choose the smoothing parameter. More details about this estimation method can be found in \citet{Chen2011}.

\subsection*{Step 2B}
Given the population mean function $\mu(t)$, fixed coefficients $\alpha$ and time-varying coefficients $\beta(t)$, i.e., given the value of $\beta$, obtain the estimates of the covariance functions of the random effects and the variance functions of the measurement errors of these two data sets to update the matrices $S_{k}$ and $M_{k}$. Specifically, model (1) can be re-written as follows, and this model is used to analyze the two sets of data separately:
\begin{equation}\label{g4}
y_{ij}^{*}=\nu_{i}(t_{ij})+\epsilon_{ij}(t_{ij}),
\end{equation}
\[\nu_{i}(t)\sim W(0,\gamma),\ \epsilon_{i}\sim N(0,V_{i}^{\frac{1}{2}}R_{i}(\theta)V_{i}^{\frac{1}{2}}),\]
\[ V_{i}=diag\{\sigma^{2}(t_{i1}),\dots,\sigma^{2}(t_{i,mi})\},\]
\[i=1,\dots,n,\ j=1,\dots,m_{i},\]
where $y_{ij}^{*}=y_{ij}-\mu(t_{ij})-x_{ij}^T\alpha-\omega_{ij}\beta(t_{ij})$. Let
$Y_{i}^{*}=(y_{ij}^{*})_{j=1,\dots,m_{i}}$, $X_{i}=(x_{i},B_{\mu}^{i},B_{c}^{i}),$
$\beta=(\alpha^{T},\beta_{\mu}^{T},\beta_{c}^{T})^{T}$,
$Z_{i}=(B_{\nu}^{T}(t_{i1}),\dots,B_{\nu}^{T}(t_{im_{i}}))^{T},$ and
$B_{c}^{i}=(\omega_{i1}B_{c}^{T}(t_{i1}),\dots,\omega_{im_{i}}B_{c}^{T}(t_{im_{i}}))^{T}.$

Define the penalized joint log likelihood of $Y_{i}^{*}$ and $\xi_{i}$ as follows:
\[
\sum_{i=1}^{n}\{(Y_{i}^{*}-Z_{i}\xi_{i})^{T}(V_{i}^{\frac{1}{2}}R_{i}V_{i}^{\frac{1}{2}})^{-1}
(Y_{i}^{*}-Z_{i}\xi_{i})+\xi_{i}^{T}\Omega^{-1}\xi_{i}\}
\]
\begin{equation}
+\lambda_{\mu}\beta_{\mu}^{T}P_{\mu}\beta_{\mu}+\lambda_{c}\beta_{c}^{T}P_{c}\beta_{c}
+\lambda_{\eta}\eta^{T}P_{\eta}\eta+\lambda_{\nu}\sum_{i=1}^{n}\xi_{i}^{T}P_{\nu}\xi_{i},\label{g5}
\end{equation}
where $\lambda_{\mu}$, $\lambda_{c}$, $\lambda_{\nu}, $ and $\lambda_{\eta}$ are smoothing parameters and $P_{\mu}$, $P_{c}$, $P_{\nu}$, and $P_{\eta}$ are penalty matrices depending on the chosen basis.
For example, if we choose the $p$th-order truncated polynomial basis functions with $S$ knots, the penalty matrix is $diag(0_{p+1},1_{S})$. Similar penalty was used in \citet{Wu2006penality},
\citet{chen2018semiparametric}, \citet{Krafty2008}, \citet{chen2018pseudo} and \citet{chen2021instrument} for smoothing splines.

Given the values of the variance components $\Omega$, $V_{i}$, and $R_{i}$,
minimize the joint penalized likelihood model (5) with respect to $\xi_{i}$ to obtain
\begin{equation}\label{g6}
\hat{\xi_{i}}=\hat{\Omega}_{\lambda_{\nu}}^{*}Z_{i}^{T}\hat{\Sigma}_{i}^{-1}Y_{i}^{*},
\end{equation}
where $\hat{\Sigma}
_{i}=Z_{i}\Omega_{\lambda_{\nu}}^{*}Z_{i}^{T}+V_{i}^{\frac{1}{2}}R_{i}V_{i}^{\frac{1}{2}}$,
$\hat{\Omega}_{\lambda_{\nu}}^{*}=(\hat{\Omega}^{-1}+\lambda_{\nu}P_{\nu})^{-1}$, and we can get the
estimates of the between-subject variance components $\Omega$ through restricted maximum likelihood
\begin{equation}\label{g7}
\hat{\Omega}=\frac{1}{n}\sum_{i=1}^{n}\{\hat{\xi_{i}}\hat{\xi_{i}}^{T}+\hat{\Omega}_{\lambda_{\nu}}^{*}
-\hat{\Omega}_{\lambda_{\nu}}^{*}Z_{i}^{T}M_{i}Z_{i}\hat{\Omega}_{\lambda_{\nu}}^{*}\},
\end{equation}
where $M_{i}=\hat{\Sigma}_{i}^{-1}-\hat{\Sigma}_{i}^{-1}X_{i}(\sum_{i=1}^{n}X_{i}^{T}
\hat{\Sigma}_{i}^{-1}X_{i}+P_{\lambda_{\nu},\lambda_{c}})^{-1}X_{i}\hat{\Sigma}_{i}^{-1},$
and $P_{\lambda_{\nu},\lambda_{c}}=diag(0_{p_{x}},\lambda_{\mu}P_{\mu},\lambda_{c}P_{c})$,
$p_{x}$ is the column dimension of $X_{i}$.

The estimation process can be summarized as follows. First let $\Omega_{0}=diag\{1,\dots,1\},\lambda_{\nu}=1,$ $\Omega_{(0)}^{*}=(\Omega_{(0)}^{-1}+\lambda_{\nu}P_{\nu})^{-1}$,
$\hat{\xi}_{i(0)}=\Omega_{(0)}^{*}Z_{i}^{T}\hat{\Sigma}_{i(0)}^{-1}Y_{i}^{*}$,
then repeat the following two steps (Step 2B(i) and Step 2B(ii)) until convergence is reached.
Step 2B(i): A Newton-Raphson algorithm based method is applied to estimate $\theta$ and $\eta$.
Step 2B(ii): One calculates the estimates of $\xi$ and $\Omega$ based on expressions (6) and (7). A likelihood-based selection approach to choose the smoothing parameter is employed. A similar estimation method can be found in \citet{Chen2011}.

When the iteration between Step 2B(i) and Step 2B(ii) stops, we obtain the estimates of the covariance functions $(\gamma_{k}(t,s)$, $k=1,2)$ associated with the random effects and the variance functions $(\sigma_{k}^{2}(t)$, $k=1,2)$ associated with errors, based on which we can estimate $S_{k}$ and $M_{k}$. Denote $\Omega^{(k)}=\mathrm{cov}(S_{k}\xi_{i})$. Since $\gamma_{1}(t,s)=B_{\nu}(t)\Omega^{(1)}B_{\nu}^{T}(s)$,
$\gamma_{2}(t,s)=B_{\nu}(t)\Omega^{(2)}B_{\nu}^{T}(s)$, let $S_{1}=I_{p+1+q}$, where $p$ is the order of spline basis function, and $q$ is the number of knots, and then $S_{2}\Omega^{(1)}S_{2}^{T}=\Omega^{(2)}$, the matrix $S_{2}$ can be obtained by the Cholesky decomposition of $\Omega^{(1)}$ and $\Omega^{(2)}$. Similarly, let $M_{1}=I_{p+1+q}$,
$c(t)=\frac{log\,\sigma_{2}^{2}(t)}{log\,\sigma_{1}^{2}(t)}$, then $B_{\sigma}^{(2)}(t)=B_{\sigma}(t)M_{2}=c(t)B_{\sigma}(t)$. In this way, the estimates of $S_{k}$ and $M_{k}$ are updated and Step 2B is completed.

Repeat Step 2A and Step 2B until convergence is reached.

\section{Simulation Study}
In this section, the performance of the proposed method are investigated through simulations. Here is a brief summary about the simulation process. First, two sets of data are generated according to a functional mixed effects model, the only difference between them are the covariance functions associated with random effects and the variance functions associated with errors. Then the difference is expanded in various ways to produce more data sets. The proposed method and the method used in \citet{Chen2011} and \citet{chen2018functional} are applied to the simulated data and their results are compared.

\subsection{Data generation}
Four cases: I, II, III, IV are considered in generating the simulation data. In Case I, two sets of data are generated from the following model:
\begin{equation}\label{g8}
y_{ij}^{(k)}=\mu(t_{ij})+(x_{ij}^{(k)})^T\alpha+\omega_{ij}^{(k)}\beta(t_{ij})
+b_{i0}^{(k)}+b_{i1}^{(k)}\cdot \nu(t_{ij})+\epsilon^{(k)}_{ij}(t_{ij}),
\end{equation}
where $k=1,2$ represents the index of the two sets. When $k=1$, the parameters are specified as follows:
\[\alpha =(1,0.02,0.02)^{T},\;\;t\in [0,1],\]
\[\mu(t)=0.5\sin(2\pi t),\;\;\beta(t)=\sqrt{\frac 1 2 t},\]
\[\nu(t)=\exp\left\lbrace-10(t-0.5)^2 \right\rbrace,\;\;\sigma_{1}^2(t)=\exp(t).\]
Note that $b_{i0}^{(1)}+b_{i1}^{(1)}\cdot \nu(t_{ij})$ is the functional random effect, the random coefficients $b_{i0}^{(1)}$ and $b_{i1}^{(1)}$ are generated from $N(0,2)$ and $N(0,1)$, respectively, and these determine the covariance function of the random effect. The three components of vector $x_{ij}^{(1)}$ are generated from $U(-1,1)$, $U(-10,10)$, and $U(-20,20)$, respectively and they are independent of each other. $\omega_{i}^{(1)}$s
are generated from Bernoulli distribution with probability 0.6. The errors $\epsilon^ {(1)}_{ij}(t_{ij})$ are independently generated from Gaussian processes with variance function $\sigma_{1}^2(t)$. The total number of subject is $n_{1}=30$, the number of observations per subject is $m=10$, and the observation time points are generated from $U(0,1)$.

When $k=2$, most of the set up are the same except that $\sigma_{2}^2(t)=4\exp(t)$, and the random coefficients $b_{i0}^{(2)}$ and $b_{i1}^{(2)}$ are generated from $N(0,8)$ and $N(0,4)$, respectively. These indicate that the covariance function of the random effects and the variance function of the errors are both larger than those when $k=1$. A combined data set consists of these two data sets, and 200 sets of combined data are generated.

To expand the difference between the two data sets within the combined data, Cases II, III and IV are considered. In Case II, the random coefficients for the second data set $b_{i0}^{(2)}$ and $b_{i1}^{(2)}$ are generated from $N(0,16)$ and $N(0,8)$, respectively, i.e., the difference between the covariance functions of the two data sets are enlarged; in Case III, the variance function for the second data set $\sigma_{2}^2(t)=16\exp(t)$, i.e., the difference between the variance functions of the two data sets are enlarged; in Case IV, the random coefficients $b_{i0}^{(2)}$
and $b_{i1}^{(2)}$ are generated from $N(0,16)$ and $N(0,8)$, respectively, and $\sigma_{2}^2(t)=16\exp(t)$, i.e., both the difference in the covariance functions and the variance functions between the two data sets are enlarged. All the other settings are the same as in Case I.

\subsection{Simulation results}
Since the main focus of this research is on the estimation of the unknown functions $\mu(t)$ and $\beta(t)$, the performance of the methods is evaluated by the confidence bands and the average mean square errors ($AMSEs$) of $\mu(t)$ and $\beta(t)$. Indeed, for each combined dataset, $\mu(t)$ and $\beta(t)$ are estimated at the time points $\{0.05,0.06,\dots,0.95\}$, and the mean square error $(MSE)$ is obtained by averaging the squared errors over 200 runs at each time point. The $MSEs$ are then averaged over all the time points to obtain the $AMSE$. For the fixed coefficient $\alpha$, the $MSEs$ of its three components can be obtained, their average value gives the $AMSE$ of $\alpha$.

The proposed method (NEW) and the method used in \citet{Chen2011} (CW) are applied to the simulated data and the results are shown in Figures 2-5 and Tables 1-2. It is clearly observed that in all the cases, the $95\%$ confidence bands of the estimated functions obtained by method NEW is narrower than those obtained by method CW. In particular, when the difference between the covariance functions of the random effects and the variance functions of the errors are enlarged between the two data sets as in Cases II, III, and IV, the confidence bands of the estimated functions obtained by method NEW do not change much, but those obtained by method CW become much wider.

Similarly, it can be observed from Table 1 that the $AMSEs$ of fixed coefficient $\alpha$, population mean function $\mu(t)$, and time-varying coefficients $\beta(t)$ obtained by method NEW are smaller than those obtained by method CW. In particular, when the difference between the covariance functions of the random effects and the variance functions of the errors are enlarged between the two data sets as in Cases II, III, and IV, the $AMSEs$ obtained by method NEW are nearly unchanged, or the change is relatively small, while the change of $AMSEs$ obtained by method CW is more obvious.

The same results can be seen more clearly from Table 2, where $RMSE$ which represents the ratio of $AMSE$ obtained by method CW over that obtained by method NEW is presented instead of $AMSE$. The $RMSEs$ of method NEW are all 1 by definition. Observe that the $RMSEs$ of method CW are all greater than 1, indicating that method CW has larger $AMSEs$, thus is less powerful than method NEW in these cases. In addition, as the difference between the covariance functions of the random effects and the variance functions of the errors between the two data sets increases in Cases II, III and IV, the $RMSEs$ of method CW become even larger.

In conclusion, the proposed method NEW behaves much better than the existing method CW both in terms of confidence bands and the $AMSEs$.

\section{Real data analysis}
The infant growth data introduced in section 2 are analyzed. To cluster the observations in the cross-sectional data set, we first divided the all the observations into 16 groups according to gender and province, since these are the two major factors of child's growth due to a large number of research. Then K-means clustering was applied to each group based on three continuous variables: the infant's birth weight and parental heights. These variables are chosen based on the application of the model in \citet{Chen2011} to the real longitudinal data with a stepwise variable selection procedure. Discussion with doctors  confirms the appropriateness of using these variables. Since it would be better if the number of observations for each pseudo individual is as close as possible to the number of observations for the real individuals, we chose to cluster the data into 197 groups (pseudo individuals), which forms a pseudo longitudinal data set.

The pseudo longitudinal data and the real longitudinal data are combined and analyzed by both the proposed method NEW and the existing method CW. The results are also compared to those obtained from a single data set, i.e., either the pseudo longitudinal data set or the real longitudinal data set. Below are the details.

\subsection{Application of method CW}
Method CW is applied to analyze the pseudo longitudinal data set, the real longitudinal data set and the combined data set, respectively. The functional mixed effects model to this problem is as follows:
\begin{center}
	$height_{ij}=\alpha_1\cdot birthweight_{ij} + \alpha_2\cdot fheight_{ij} + \alpha_3\cdot mheight_{ij}$\\
	$+\mu(t_{ij})+\beta(t_{ij})\cdot sex_{ij}+ \nu_i(t_{ij}) + \epsilon_{ij}(t_{ij}),$
\end{center}
where $height_{ij}$, $birthweight_{ij}$, $fheight_{ij}$, $mheight_{ij}$, and $sex_{ij}$ represent the height, birth weight, father's height, mother's height, and sex of baby $i$ measured at visit $j$. The value of $sex_{ij}$ is one for boy and zero for girl. $t_{ij}$ is the corresponding age, $\mu(t)$ is the mean height function, $\beta(t_{ij})$ is the height difference between boys and girls over time, $\nu_i(t)$ is the subject-specific random effect, and $\epsilon_{ij}(t)$ is the measurement error.

The estimated mean function and time-varying coefficient and their associated 95$\%$ confidence bands are shown in Figures 6-8. Figure 6 shows the results for the pseudo longitudinal data set, Figure 7 shows the results for the real longitudinal data set, and Figure 8 shows the results for the combined data set. Observe that the estimated mean functions are very similar for the pseudo and real longitudinal data, but the confidence band for the real data is slightly narrower than the pseudo data, indicating smaller variation. The estimated $\beta(t)s$ look different: the result is more curved for the pseudo data than for the real data. For the combined data set, the estimate of $\mu(t)$ looks very similar to the previous two but with slightly narrower confidence band. The estimate of $\beta(t)$ looks to be a balance between the previous two estimates, it is curved a little, with confidence band narrower than the two before. This shows that combining the data does improve the estimation of the functions.

\subsection{Application of the proposed method}
The proposed method is applied to analyze the combined data. The functional mixed effects model for the combined data set is:
\begin{center}
	$height_{ij}^{(k)}=\alpha_1\cdot birthweight_{ij}^{(k)} + \alpha_2\cdot fheight_{ij}^{(k)} + \alpha_3\cdot mheight_{ij}^{(k)}$\\
	$+\mu(t_{ij})+\beta(t_{ij})\cdot sex_{ij}^{(k)}+ \nu_i^{(k)}(t_{ij}) + \epsilon_{ij}^{(k)}(t_{ij})$
\end{center}
\[k=1:i=1,\dots,197, j=1,\dots,m_{i},\]
\[k=2:i=1,\dots,251, j=1,\dots,m_{i},\]
where $height_{ij}^{(k)}$, $birthweight_{ij}^{(k)}$, $fheight_{ij}^{(k)}$, $mheight_{ij}^{(k)}$, and $sex_{ij}^{(k)}$ are the height, birth weight, father's height, mother's height and sex of baby $i$ measured at visit $j$ in the $k$th set of data, and the value of $sex_{ij}^{(k)}$ is one for boy and zero for girl, $t_{ij}$ is the corresponding age, $\mu(t)$ is the mean function, $\beta(t_{ij})$ is the height difference between boys and girls over time, and $\nu_i^{(k)}(t)$ is the random effects in the $k$th data set,
$\epsilon_{ij}^{(k)}(t)$ is the measurement error in the $k$th data set.

In the estimation, truncated quadratic splines are used for the mean function, varying coefficient and variance functions and truncated linear splines are used for the random effect curves. The number of knots is $K=\min(M/4, 40)$, where $M$ is the number of non-overlapping time points observed for all subjects. This is proposed by \citet{ruppert2002select} and \citet{krivobokova2007a}, in which it has been proved that the actual choice of $K$ and the location of knots have little influence on the resulting penalized fit as long as $K$ is large. The estimated covariance functions of the random effects and the estimated standard deviation functions of the errors of these two data sets are shown in Figures 9. The estimated mean function and time-varying coefficient and their associated 95$\%$ confidence bands obtained through a bootstrap procedure described in \citet{huang2002varying} are shown in Figure 10.

Observe from Figure 9 that in general the estimated covariance functions $\gamma_{k}(s,t)$, $k=1,2$ and estimated standard deviation functions $\sigma_{k}(t)$, $k=1,2$ for the pseudo data are larger than those for the real data, which is reasonable and suggests that the proposed model is more appropriate to use for the combined data set. The magnitude of $\gamma_{k}(s,t)$, $k=1,2$ is much bigger than $\sigma_{k}(t)$,
$k=1,2$, indicating that the dominant variance components of the variation in infant's heights is the between-subject variation.

Observe from Figure 10 that the estimated mean function $\mu(t)$ increases almost linearly over time except for a faster increase at the beginning. The time-varying coefficient function $\beta(t)$ increases rapidly before 6 months and remains almost constant afterwards. There is slight decrease and increase between 6 and 24 months. Comparing Figure 8 and Figure 10, one can see that the widths of the confidence bands obtained from method NEW are generally narrower than those obtained from method CW. In addition, the shape of estimated $\beta(t)$ by method NEW is closer to the shape estimated from the real longitudinal data while that obtained by method CW stands in between the results of real and pseudo data, which indicates that while separating the two groups in terms of their variation, method NEW naturally puts more weight on the real longitudinal data in estimating the fixed effects since its variation is smaller.

\section{Discussion}
In this paper, a new method that performs joint analysis of longitudinal and cross-sectional growth data is proposed. There are two main innovations of the method: 1) The cross-sectional data are clustered into groups so that individuals that have similar characters are grouped together to form a pseudo individual. Then combination of the cross-sectional data and longitudinal data becomes possible since both data sets have longitudinal structures, only that the pseudo data have more variation than the real data. 2) A functional mixed effects model that allows different covaiance and variance functions is developed to fit the combined data set. Both simulation and real data analysis demonstrate the usefulness of the new method. The simulation shows that the bigger the difference in variation is, the better our method performs compared to the other method, which is consistent with the conclusion in \citet{Fan2007Analysis}: accurate estimation of a covariance function leads to efficiency gain in estimating the population mean function and fixed effects parameters.

In addition to combining longitudinal data and cross-sectional data in analysis, the proposed method can be used in other occasions when the variation are different among data sets. For example, suppose there is a longitudinal data set, it is possible that $\gamma(s,t)$ or $\sigma^{2}(t)$ are different for males and females or young and old people. In these cases one could divide the longitudinal data into several groups of data by gender or age, and then apply the proposed method to analyze the whole data set. The estimation of the fixed effects would be more accurate once the variance and covariance functions are estimated well.

\clearpage
%\bigskip
%\noindent{\large\bf Acknowledgement}\\
%The work was supported by National Natural Science Foundation of China (project number: 11771146, 11831008, 81530086), the National Social Science Foundation Key Program (17ZDA091), the 111 Project (B14019) and Program of Shanghai Subject Chief Scientist (14XD1401600). We thank Professor Yanyuan Ma for helpful discussion about the paper.

\bibliographystyle{apalike}
\bibliography{chen_chen_zhou_2019}

\clearpage
\newpage

\begin{table}[h!]
	%\scriptsize
	\centering
	\caption{Simulation results in terms of AMSE}
	\begin{tabular}{cc|ccc}
		\hline
		\multicolumn{2}{c|}{}& $AMSE_{\alpha}$ & $AMSE_{\mu}$ &  $AMSE_{\beta}$ \\
		\hline
		%\multicolumn{2}{c|}{(1)} &\multicolumn{3}{c}{$n=50,\;m=20$}\\
		Case I &  NEW
		&0.006& 0.212& 0.385\\
		& CW &0.010& 0.393& 0.799\\
		Case II & NEW
		&0.006& 0.268& 0.501\\
		%& AVAR &0.2067& 0.2291& 0.2062& 0.2292\\
		&  CW &0.015& 1.061& 1.649\\
		Case III & NEW
		&0.007 &0.227&0.420 \\
		%& AVAR &0.0233 &0.0434&   -    &-\\
		&  CW &0.026 &0.629&1.063  \\
		Case IV & NEW
		&0.007 &0.237&0.394 \\
		%& AVAR &0.0233 &0.0434&   -    &-\\
		&  CW &0.032 &1.271&1.965  \\
		\hline			
	\end{tabular}
	%\end{table}
	\vspace{6mm}
	%\begin{table}[htbp]
	%\scriptsize
	\centering
	\caption{Simulation results in terms of RMSE}
	\begin{tabular}{cc|ccc}
		\hline
		\multicolumn{2}{c|}{}& $RMSE_{\alpha}$ & $RMSE_{\mu}$ &  $RMSE_{\beta}$ \\
		\hline
		%\multicolumn{2}{c|}{(1)} &\multicolumn{3}{c}{$n=50,\;m=20$}\\
		Case I & NEW
		&1& 1&1\\
		& CW &1.667&1.853&2.075\\
		Case II & NEW
		&1& 1 & 1\\
		%& AVAR &0.2067& 0.2291& 0.2062& 0.2292\\
		& CW &2.500&3.955&3.291\\
		Case III & NEW
		&1 &1&1 \\
		%& AVAR &0.0233 &0.0434&   -    &-\\
		& CW &3.714&2.770&2.530 \\
		Case IV & NEW
		&1 &1&1 \\
		%& AVAR &0.0233 &0.0434&   -    &-\\
		& CW &4.571&5.363&4.987  \\
		\hline			
	\end{tabular}
	\vspace{5mm}
\end{table}

\begin{figure}[h!]
	\centering
	\begin{minipage}[c]{0.4\textwidth}
		\centering
		\includegraphics[width=1\textwidth]{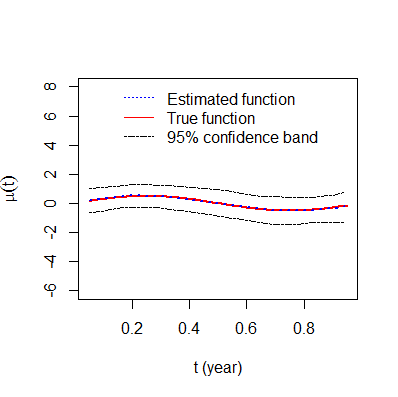}
	\end{minipage}
	\begin{minipage}[c]{0.4\textwidth}
		\centering
		\includegraphics[width=1\textwidth]{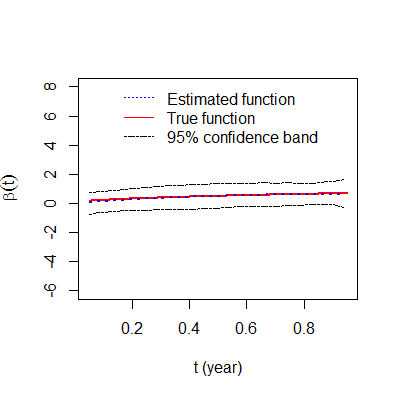}
	\end{minipage}
	\begin{minipage}[c]{0.4\textwidth}
		\centering
		\includegraphics[width=1\textwidth]{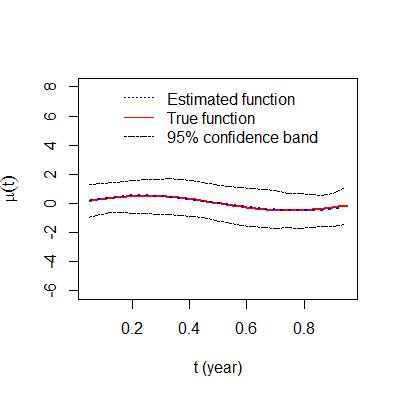}
	\end{minipage}
	\begin{minipage}[c]{0.4\textwidth}
		\centering
		\includegraphics[width=1\textwidth]{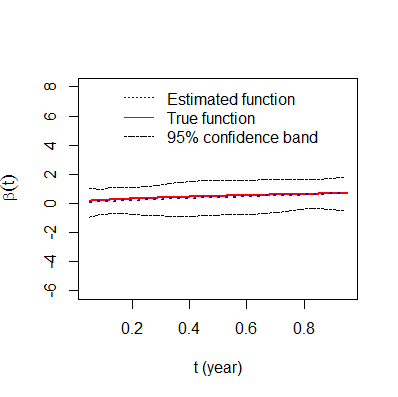}
	\end{minipage}
	\caption{\small{Estimated $\mu(t)$ and $\beta(t)$ and their $95\%$ confidence bands by method NEW (first row) and method CW (second row) in Case I.}}
	\vspace{-9mm}
\end{figure}

\begin{figure}[h!]
	\centering
	\begin{minipage}[c]{0.4\textwidth}
		\centering
		\includegraphics[width=1\textwidth]{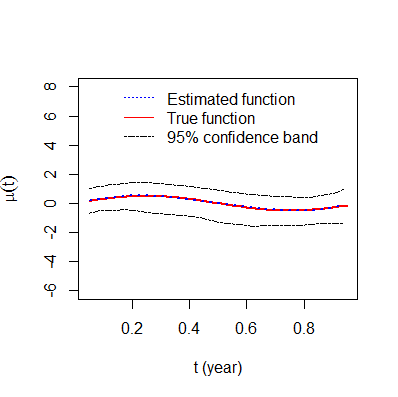}
	\end{minipage}
	\begin{minipage}[c]{0.4\textwidth}
		\centering
		\includegraphics[width=1\textwidth]{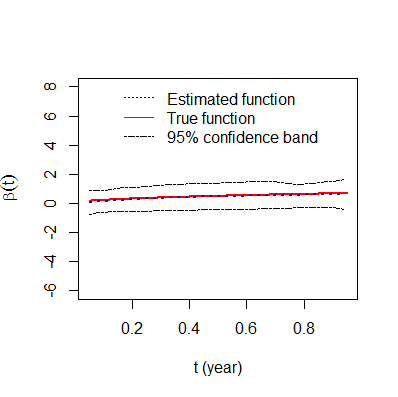}
	\end{minipage}
	\begin{minipage}[c]{0.4\textwidth}
		\centering
		\includegraphics[width=1\textwidth]{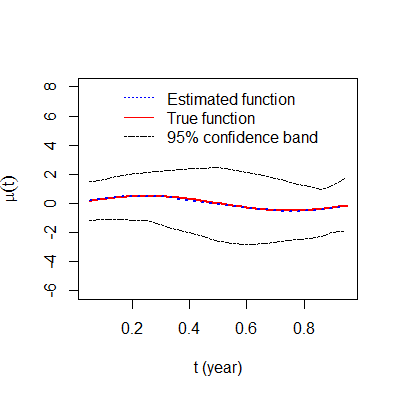}
	\end{minipage}
	\begin{minipage}[c]{0.4\textwidth}
		\centering
		\includegraphics[width=1\textwidth]{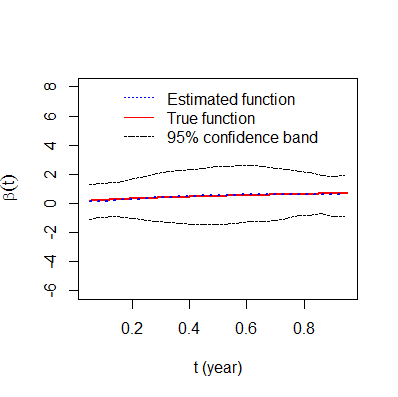}
	\end{minipage}
	\caption{\small{Estimated $\mu(t)$ and $\beta(t)$ and their $95\%$ confidence bands by method NEW (first row) and method CW (second row) in Case II.}}
	\vspace{-9mm}
\end{figure}

%\clearpage
\begin{figure}[h!]
	\centering
	\begin{minipage}[c]{0.4\textwidth}
		\centering
		\includegraphics[width=1\textwidth]{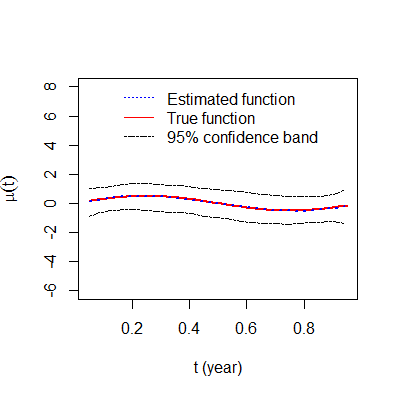}
	\end{minipage}
	\begin{minipage}[c]{0.4\textwidth}
		\centering
		\includegraphics[width=1\textwidth]{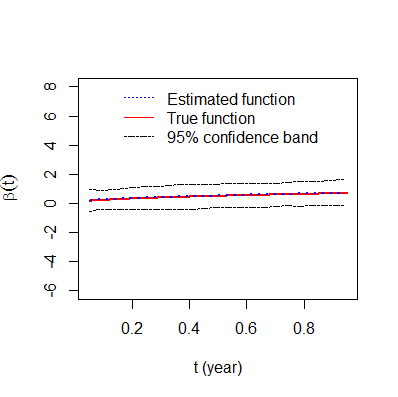}
	\end{minipage}
	\begin{minipage}[c]{0.4\textwidth}
		\centering
		\includegraphics[width=1\textwidth]{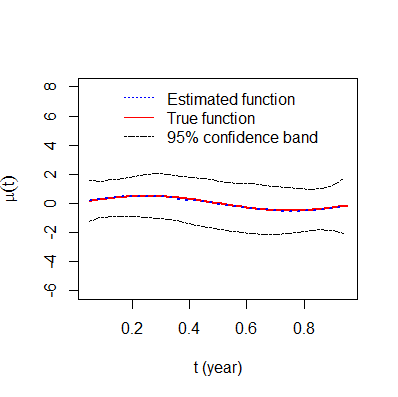}
	\end{minipage}
	\begin{minipage}[c]{0.4\textwidth}
		\centering
		\includegraphics[width=1\textwidth]{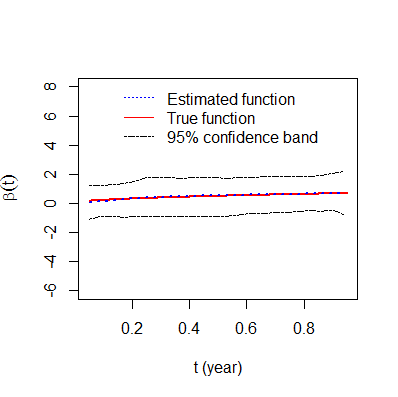}
	\end{minipage}
	\caption{\small{Estimated $\mu(t)$ and $\beta(t)$ and their $95\%$ confidence bands by method NEW (first row) and method CW (second row) in Case III.}}
	\vspace{-9mm}
\end{figure}

\begin{figure}[h!]
	\centering
	\begin{minipage}[c]{0.4\textwidth}
		\centering
		\includegraphics[width=1\textwidth]{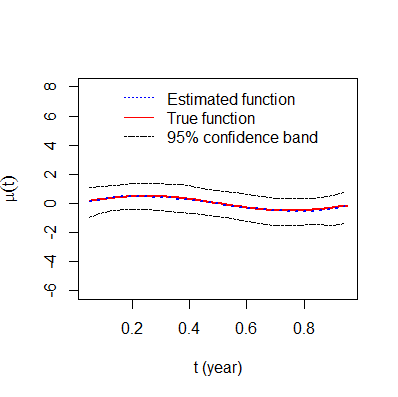}
	\end{minipage}
	\begin{minipage}[c]{0.4\textwidth}
		\centering
		\includegraphics[width=1\textwidth]{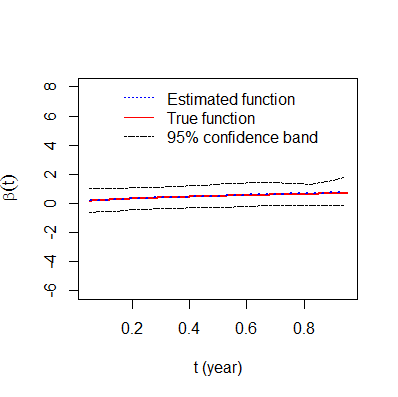}
	\end{minipage}
	\begin{minipage}[c]{0.4\textwidth}
		\centering
		\includegraphics[width=1\textwidth]{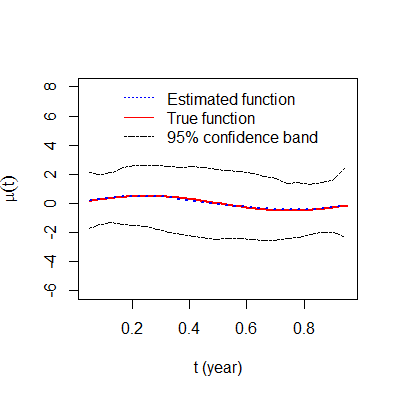}
	\end{minipage}
	\begin{minipage}[c]{0.4\textwidth}
		\centering
		\includegraphics[width=1\textwidth]{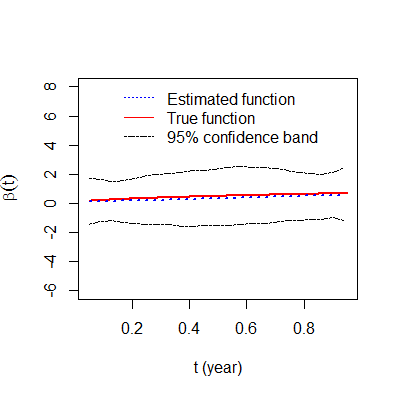}
	\end{minipage}
	\caption{\small{Estimated $\mu(t)$ and $\beta(t)$ and their $95\%$ confidence bands by method NEW (first row) and method CW (second row) in Case IV.}}
	\vspace{0mm}
\end{figure}

\clearpage
\begin{figure}[h!]
	\centering
	\begin{minipage}[c]{0.38\textwidth}
		\centering
		\includegraphics[width=1\textwidth]{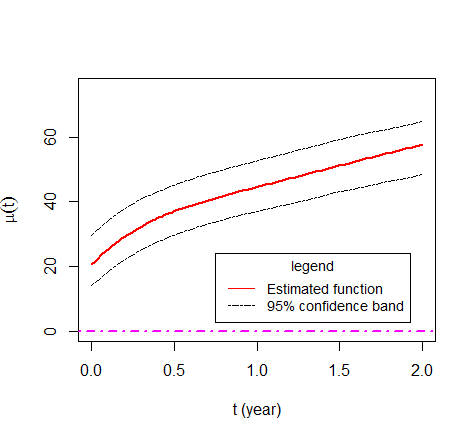}
	\end{minipage}
	\begin{minipage}[c]{0.38\textwidth}
		\centering
		\includegraphics[width=1\textwidth]{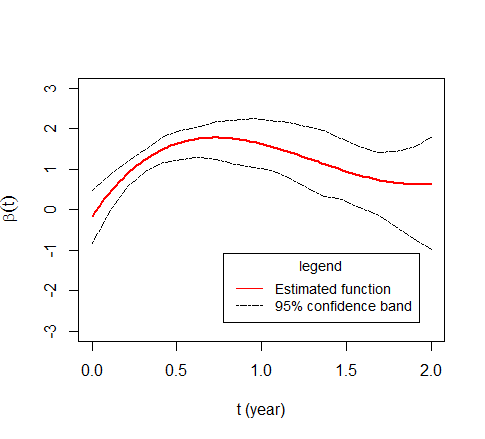}
	\end{minipage}
	\caption{\small{Estimated $\mu(t)$ and $\beta(t)$ and their $95\%$ confidence bands for the pseudo longitudinal data set by method CW.}}
	\vspace{-8mm}
\end{figure}	
\begin{figure}[h!]
	\centering
	\begin{minipage}[c]{0.38\textwidth}
		\centering
		\includegraphics[width=1\textwidth]{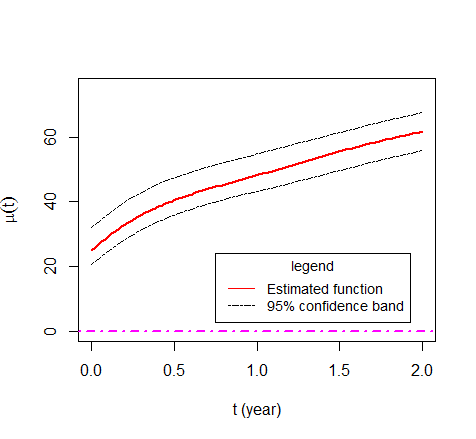}
	\end{minipage}
	\begin{minipage}[c]{0.38\textwidth}
		\centering
		\includegraphics[width=1\textwidth]{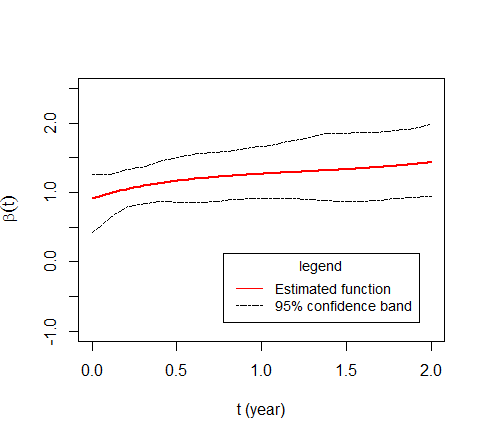}
	\end{minipage}
	\caption{\small{Estimated $\mu(t)$ and $\beta(t)$ and their $95\%$ confidence bands for the real longitudinal data set by method CW.}}
	\vspace{-8mm}
\end{figure}	
\begin{figure}[h!]
	\centering
	\begin{minipage}[c]{0.38\textwidth}
		\centering
		\includegraphics[width=1\textwidth]{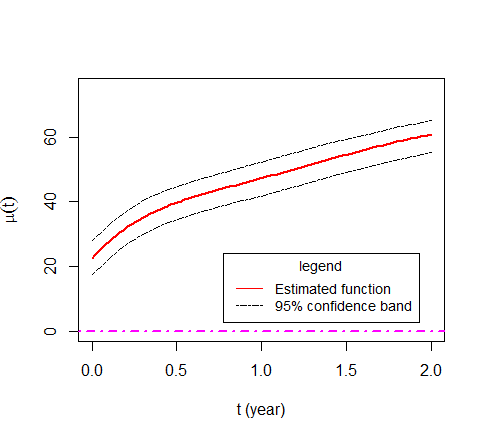}
	\end{minipage}
	\begin{minipage}[c]{0.38\textwidth}
		\centering
		\includegraphics[width=1\textwidth]{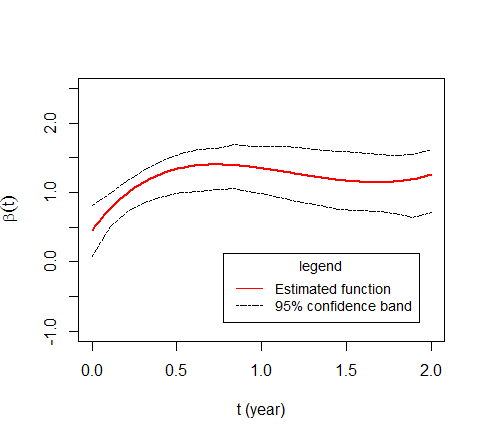}
	\end{minipage}
	\caption{\small{Estimated $\mu(t)$ and $\beta(t)$ and their $95\%$ confidence bands for the combined data set by method CW.}}
	\vspace{-8mm}
\end{figure}

\begin{figure}[h!]
	\centering
	\begin{minipage}[c]{0.42\textwidth}
		\centering
		\includegraphics[width=1\textwidth]{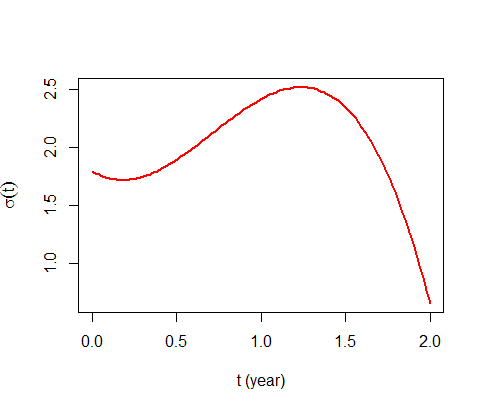}
	\end{minipage}
	\begin{minipage}[c]{0.49\textwidth}
		\centering
		\includegraphics[width=0.9\textwidth]{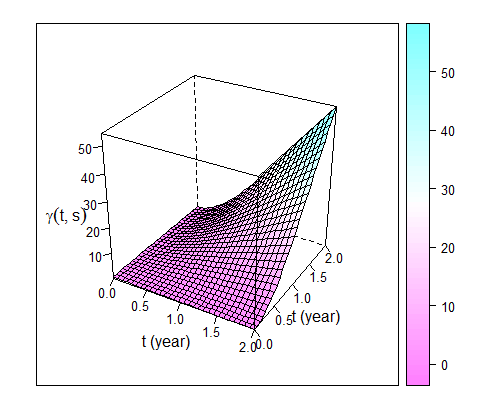}
	\end{minipage}
	\begin{minipage}[c]{0.42\textwidth}
		\centering
		\includegraphics[width=1\textwidth]{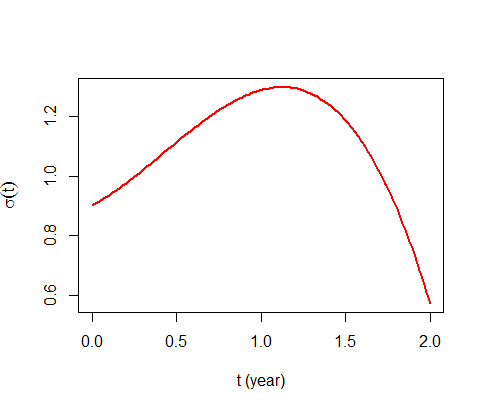}
	\end{minipage}
	\begin{minipage}[c]{0.49\textwidth}
		\centering
		\includegraphics[width=0.9\textwidth]{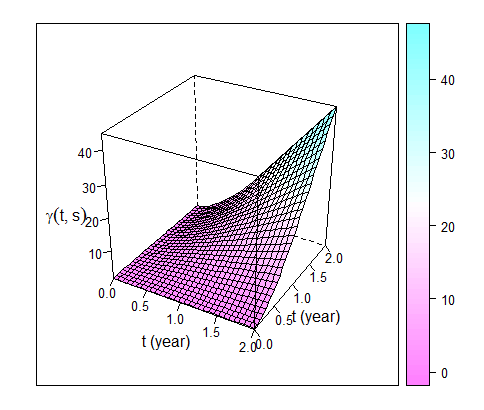}
	\end{minipage}
	\caption{\small{Estimated $\sigma(t)$ and $\gamma(t,s)$ for the pseudo longitudinal data (first row) and real longitudinal data set (second row) by method NEW.}}
	\vspace{1mm}
\end{figure}

\newpage
\begin{figure}[h!]
	\centering
	\begin{minipage}[c]{0.38\textwidth}
		\centering
		\includegraphics[width=1\textwidth]{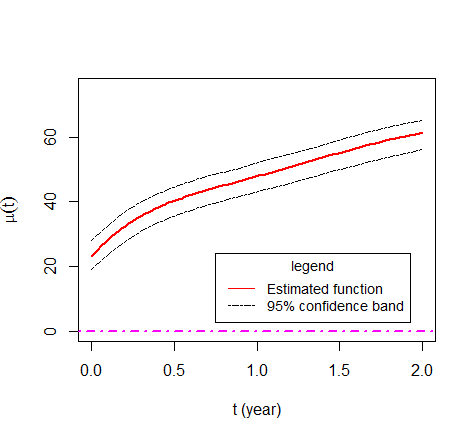}
	\end{minipage}
	\begin{minipage}[c]{0.38\textwidth}
		\centering
		\includegraphics[width=1\textwidth]{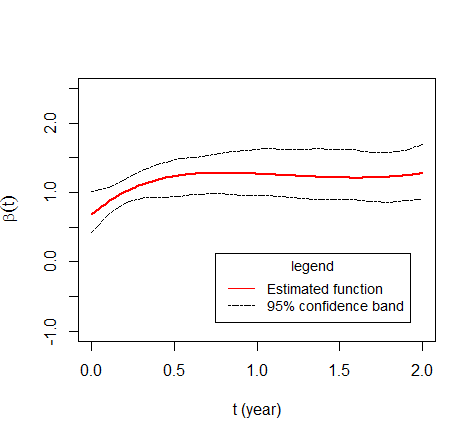}
	\end{minipage}
	\caption{\small{Estimated $\mu(t)$ and $\beta(t)$ and their $95\%$ confidence bands for the combined data set by method NEW.}}
	\vspace{-2mm}
\end{figure}

\label{lastpage}

\end{document}